\documentclass[preprint,article,12pt]{elsarticle}

\makeatletter
\def\ps@pprintTitle{%
 \let\@oddhead\@empty
 \let\@evenhead\@empty
 \def\@oddfoot{}%
 \let\@evenfoot\@oddfoot}
\makeatother

\usepackage{natbib}
\usepackage{amssymb}
\usepackage{amsthm}
\usepackage{amsmath}
\usepackage{hyperref}
\usepackage{algorithm}
\usepackage{algpseudocode}
\usepackage{ctable}
\newcolumntype{K}{>{\centering\arraybackslash}X}

\newcommand{\vect}[1]{\mathbf{#1}}
\newcommand{\vers}[1]{\hat{\mathbf{#1}}}

\begin{document}

\begin{frontmatter}

\title{Fast overlap detection between hard-core colloidal cuboids and spheres. The OCSI algorithm}

\author[lab1]{Luca Tonti}
\author[lab1]{Alessandro Patti\corref{*}}
\cortext[*]{E-mail: alessandro.patti@manchester.ac.uk}
\address[lab1]{Department of Chemical Engineering and Analytical Science, The University of Manchester, Manchester, M13 9PL, UK}


\begin{abstract}
Collision between rigid three-dimensional objects is a very common modelling problem in a wide spectrum of scientific disciplines, including Computer Science and Physics. It spans from realistic animation of polyhedral shapes for computer vision to the description of thermodynamic and dynamic properties in simple and complex fluids. For instance, colloidal particles of especially exotic shapes are commonly modelled as hard-core objects, whose collision test is key to correctly determine their phase and aggregation behaviour. In this work, we propose the OpenMP Cuboid Sphere Intersection (OCSI) algorithm to detect collisions between prolate or oblate cuboids and spheres. We investigate OCSI's performance by bench-marking it against a number of algorithms commonly employed in computer graphics and colloidal science: Quick Rejection First (QRI), Quick Rejection Intertwined (QRF) and SIMD Streaming Extensions (SSE). We observed that QRI and QRF significantly depend on the specific cuboid anisotropy and sphere radius, while SSE and OCSI maintain their speed independently of the objects' geometry. While OCSI and SSE, both based on SIMD parallelization, show excellent and very similar performance, the former provides a more accessible coding and user-friendly implementation as it exploits OpenMP directives for automatic vectorization. 
\end{abstract}

\end{frontmatter}


\section{Introduction}
\label{intro}
Employing computer programs and algorithms to generate 2D or 3D images is referred to as rendering. Rendering is a topic of striking relevance in computer graphics with practical impact on many heterogeneous disciplines, spanning engineering, simulators, video games and movie special effects. Collision detection and collision determination are key elements of rendering as they determine the distance between two objects and their possible intersection \cite{Ericson2004}. Due to their widespread use in video representation of time-evolving systems, with tens of frames displayed per second, algorithms for rendering are expected to be very efficient \cite{RTR4,Chang2010}. Generally, to assess whether two complex objects collide, the distance between their respective bounding volumes is evaluated first. Common bounding volumes are cuboidal boxes, whose axes might or might not be aligned, or spheres. Due to their simple geometry, the collision between cuboids and/or spheres is computationally easier \cite{Gottschalk1996, Arvo1990, Ratschek1994, Larsson2007}, thus enhancing the speed and efficiency of the overall rendering process \cite{RTR4}. Collision detection algorithms are of utmost relevance in many heterogeneous applications spanning computer graphics for shape modelling and video games \cite{Moore1988,Pungotra2008}, robotics to prevent potential collisions in man–robot interactions \cite{Yang2017} or machining of sculptured surfaces \cite{Tang2014}, and simulations of molecular or particle systems to estimate their thermodynamics properties \cite{Frenkel1988, ANDERSON2020109363}. 

Collision algorithms have also been key to address the thermodynamics of liquid and solid phases and their phase transition by early molecular simulation studies that employed the hard-sphere model \cite{Rosenbluth1954, Wood1957, Alder1957}. More recently, and following the seminal theory by Onsager on the isotropic-to-nematic transition of hard rods \cite{Onsager1949}, they were fundamental to confirm the crucial role of excluded volume effects in the formation of  colloidal liquid crystal phases of anisotropic particles \cite{Frenkel1988}. Realising the practical impact of the particle shape on the design of nanomaterials triggered the blooming of biosynthetic \cite{shankar2004}, chemical \cite{sun2002} and  physical \cite{manoharan2003} experimental routes to manufacture precise building blocks with \textit{ad hoc} properties, including lock-and-key particles \cite{sacanna2010}, fused spheres \cite{sacanna2013}, superballs \cite{rossi2015} and cuboids \cite{xiang2006, okuno2010, cortie2012, khlebtsov2015}. The appearance of these exotic shapes unveiled a realm of novel opportunities in nanomaterials science by offering an increasingly varied selection of morphologies for state-of-the-art applications spanning medicine (controlled drug delivery), smart materials (self-healing coatings) and photonics (light detection), among others. Often anticipating experimental evidence, computer simulations have significantly contributed to our comprehension of the effect of particle shape and interaction at the nanoscale on the material properties at the macroscale \cite{glotzer2007, damasceno2012, vananders2014, denijs2014}. Understanding the fundamentals of such a complex correlation, which develops over orders of magnitude in length and time scales,  dramatically depends on the existence of reliable force fields mimicking the interactions between particles. This is not always the case for most exotic particle shapes, whose force field is assumed to be described by mere excluded volume effects and thus only incorporates a hard-core interaction potential. Consequently, efficient and robust algorithms able to detect collisions and intersections between objects become essential to extract structural, thermodynamic and dynamic properties of such systems from a molecular simulation. In colloid science, cuboids are especially intriguing building blocks that can form a rich variety of liquid crystal phases \cite{cuetos2017, patti2018, cuetos2019, cuetos2020, effran2020}. Incorporating guest spherical particles in these phases is relevant to understand phenomena of diffusion in crowded environments that display a significant degree of ordering. 

In the light of these considerations, highlighting the harmonious inter-disciplinary convergence of computer graphics and colloid science, here we report on the specific case of cuboid-sphere collision detection. In particular, we propose our own OpenMP \cite{Dagum1998} Cuboid Sphere Intersection (OCSI) algorithm to detect collisions in monodisperse systems of cuboids and spheres oriented in a 3D space. OCSI is found to be especially efficient when compared to the Quick Rejection First (QRI) and the Quick Rejection Intertwined (QRF) algorithms, and more user-friendly and easier to implement than the SIMD Streaming Extensions (SSE) algorithm. This paper is organised as follows. In Section 2, we  detail the theoretical framework of the cuboid-sphere intersection problem and the state-of-the-art in software implementation. In Section 3, we describe the code that we have specifically developed to test each of the above-mentioned algorithms' efficiency for cuboids of different shape and spheres of different size. The comparison between the algorithms is then discussed in Section 4, while, in Section 5, we draw our conclusions. 


\section{Algorithms}
\label{method}

In geometry, a sphere $\mathcal{S}$ is identified by its radius, $R$, and the position of its centre, $\vect{{r}_S}$, with respect to a reference point. Similarly, a cuboid $\mathcal{C}$ can be defined by the extension of its thickness, $2c_{T}$, length, $2c_{L}$, and width, $2c_{W}$, the position of its centre of mass, $\vect{{r}_C}$, and the unit vectors $\vers{e}_\vect{T}$, $\vers{e}_\vect{L}$ and $\vers{e}_\vect{W}$ that indicate the orientation of its three orthogonal axes. As a result, all the points within the volume occupied by the cuboid can be indicated by a vector $\vect{C}$ that reads

\begin{equation}
\vect{C}=\vect{r_C}+\sum_{i=T,L,W} \alpha_{i}c_{i}\vers{e}_\vect{i},
\label{eq:def_cub2}
\end{equation}

\noindent where $T$, $L$ and $W$ indicate, respectively, the cuboid thickness, length and width, whereas $\alpha = \big[-1,1\big]$ is a scalar interval. With these essential definitions, the minimum distance, $d_{min}$, between the surface of a randomly oriented cuboid and the center of a sphere can be calculated as follows:

\begin{equation}
d_{min}=\sqrt{ \sum_{i=T,L,W} \Theta \Big(\big|\vect{r_{SC}} \cdot \vers{e}_\vect{i}\big| -c_{i}  \Big) \Big\{ \big|\vect{r_{SC}} \cdot \vers{e}_\vect{i}\big| -c_{i}  \Big\}^2} ,
\label{mindist2}
\end{equation}

\noindent where $\vect{r_{SC}} = \vect{r_S} - \vect{r_C}$ and $\Theta$ is the Heaviside step function:

\begin{equation}
\Theta(x) = \bigg\{
\begin{matrix}
0 & x \le 0  \\
1 & x > 0 
\end{matrix}
\end{equation}

\noindent The interested reader is referred to \ref{app:obbsphmin} for a formal derivation of Eq.\ \ref{mindist2}. To the best of our knowledge, Arvo was the first to propose an algorithm detecting the intersection between a sphere and an axis-aligned cuboid, that is a cuboid whose orientation matches that of the reference axes \cite{Arvo1990}. For this specific case, we assume that the cuboid thickness is aligned with the $x$ axis, i.e. $\vers{e}_\vect{T}=\vers{x}$, its length with the $y$ axis, i.e. $\vers{e}_\vect{L}=\vers{y}$, and its width with the $z$ axis, i.e. $\vers{e}_\vect{W}=\vers{z}$. Following this assumption, Eq.\ \ref{eq:def_cub2} can be rewritten as 

\begin{equation}
\begin{gathered}
\vect{C}=\vect{r_C}+\alpha_{T}c_{T}\vers{x}+\alpha_{L}c_{L}\vers{y}+\alpha_{W}c_{W}\vers{z} = \\
=\vect{r_C}+\big[-c_{T},c_{T}\big]\vers{x}+\big[-c_{L},c_{L}\big]\vers{y}+\big[-c_{W},c_{W}\big]\vers{z} = \\
=\big[r_{C,x}-c_{T},r_{C,x}+c_{T}\big]\vers{x}+\big[r_{C,y}-c_{L},r_{C,y}+c_{L}\big]\vers{y}+ \\ 
+\big[r_{C,z}-c_{W},r_{C,z}+c_{W}]\vers{z} =\\ =\sum_{i=x,y,z}B_{i}\vers{i}
\end{gathered}
\end{equation}

\noindent where $\vers{i}=\vers{x}, \vers{y}, \vers{z}$ are the reference axes for $T$, $L$ and $W$, respectively, and $B_{x}=\big[r_{C,x}-c_{T},r_{C,x}+c_{T}\big]$, $B_{y}=\big[r_{C,y}-c_{L},r_{C,y}+c_{L}\big]$ and $B_{z}=\big[r_{C,z}-c_{W},r_{C,z}+c_{W}\big]$.
Therefore, for an axis-aligned cuboid, $d_{min}$ can be calculated as 

\begin{equation}
d_{min}= \sqrt{\sum_{i=x,y,z} \Big\{\min \big(r_{S,i}-B_{i} \big)\Big\}^2}.
\label{mindist_algn}
\end{equation}

\noindent By using the infimum and supremum of $B_{i}$, the terms in the above sum can be easily calculated:

\begin{enumerate}
    \item if $r_{S,i} < B_{i,inf}$, then $\min \big(r_{S,i}-B_{i} \big) = B_{i,inf} - r_{S,i}$,
    \item if $r_{S,i} > B_{i,sup}$, then $\min \big(r_{S,i}-B_{i} \big) = r_{S,i} - B_{i,sup}$, 
    \item if $r_{S,i} \in B_{i}$, then $\min \big(r_{S,i}-B_{i} \big) = 0$. 
\end{enumerate}

\noindent Consequently, the algorithm proposed by Arvo only requires the extreme values of $B_{x},B_{y},B_{z}$ along with the sphere radius and position and detects cuboid-sphere collisions if $d_{min} \leq R$. An illustrative example of a pseudocode describing its main steps is reported in Algorithm \ref{alg:arvo}.

\begin{algorithm}[ht!]
\caption{- Arvo}
\begin{algorithmic}[1]
\Function{Arvo}{$\vect{r_{S}}, R ,B_{i,inf}, B_{i,sup}$}
\State ${d} \gets 0$ \Comment{initialising minimum distance}
\For{$i \in \left\{x,y,z \right\} $}
    \If{($r_{S,i} < B_{i,inf}$)}
        \State $d \gets d + \left(B_{i,inf} - r_{S,i} \right)^2$
    \ElsIf{($r_{S,i} > B_{i,sup}$)}
        \State $d \gets d + \left(r_{S,i} -B_{i,sup}\right)^2$
    \EndIf
\EndFor
\State \textbf{if} ($d \leq R^2$) \textbf{return} $true$ \Comment{checking overlap}
\State \textbf{return} $false$
\EndFunction
\end{algorithmic}
\label{alg:arvo}
\end{algorithm}

The alignment of the cuboid unit vectors with the reference axes is a particular case of a more general scenario with the cuboid randomly oriented. Eventually, Arvo's algorithm can also be applied to randomly oriented cuboids by performing a translation of the vectors involved in the calculation of $d_{min}$ in the reference frame of $\mathcal{C}$. Rokne and Ratschek proposed to estimate $d_{min}$ by employing interval analysis and reported a test to determine whether a point $P \in \mathcal{C}$ is within a sphere delimited by four peripheral points ~\cite{Ratschek1994}. The algorithms proposed by Larsson and co-workers employ quick rejection and vectorized overlap tests to enhance the efficiency of collision detection between a sphere and either an aligned or a randomly oriented cuboid \cite{Larsson2007}. The pseudocode of these algorithms, referred to as Quick Reference Intertwined (QRI) and Quick Reference First (QRF), are reported in Algorithm \ref{alg:qri} and Algorithm \ref{alg:qrf}, respectively. Both QRI and QRF are based on the implementation of a quick rejection test that immediately excludes an overlap if at least one of the summands in Eq. \ref{mindist2} or Eq. \ref{mindist_algn} is larger than $R^2$. For the general case of a randomly oriented cuboid, this condition reads

\begin{equation}
\begin{gathered}
\Big\{ \big|\vect{r_{SC}} \cdot \vers{e}_\vect{i}\big| -c_{i}  \Big\}^2 > R^2 \; \Leftrightarrow\\
\vect{r_{SC}} \cdot \vers{e}_\vect{i} < -c_{i} -R \; \cup \; \vect{r_{SC}} \cdot \vers{e}_\vect{i} > c_{i} + R. \\
\forall \; i=T,L,W
\label{eq:spec_case}
\end{gathered}
\end{equation}

\begin{figure}[ht!]
    \centering
    \includegraphics{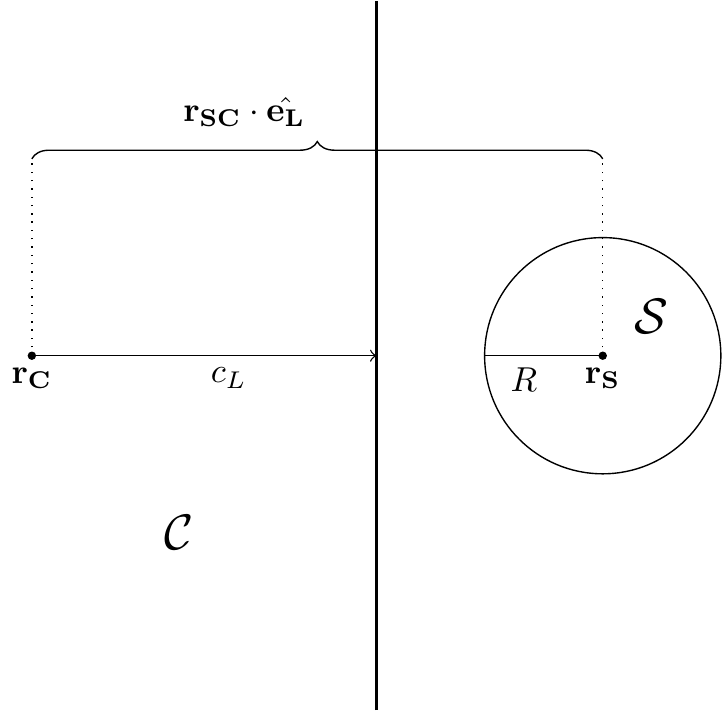}
    \caption{Schematic representation of a sphere $\mathcal{S}$ and a cuboid $\mathcal{C}$ at relative distance $\rm \vect{r_{SC}} \cdot \vers{e}_L$. Sphere and cuboid are centered, respectively, at $\vect{r_S}$ and $\vect{r_C}$, and $c_L$ is half of the cuboid length. If $\vect{r_{SC}} \cdot \vers{e}_L > c_L + R$, then $\mathcal{S}$ and $\mathcal{C}$ do not overlap.}
    \label{fig:reject}
\end{figure}

\noindent A geometrical representation of this condition is provided in Fig. \ref{fig:reject}, where a sphere $\mathcal{S}$ of radius $R$ and centered at $\vect{r_S}$ is at the distance $\vect{r_{SC}} \cdot \vers{e}_\vect{L}$ from the center of mass of a cuboid $\mathcal{C}$ that is centered at $\vect{r_C}$. For this specific arrangement, the left-hand side of Eq. \ref{eq:spec_case} measures the distance of $\mathcal{S}$ from the face of $\mathcal{C}$ delimited by $T$ and $W$ and schematically identified by the vertical solid line of Fig. \ref{fig:reject}. 

\noindent QRI applies this rejection criterion within the loop that evaluates the minimum distance, precisely at lines \ref{qri:out1} and \ref{qri:out2} of Algorithm \ref{alg:qri}. By contrast, QRF performs the three quick rejection tests, one for each summand of Eq. \ref{mindist2}, before the computation of the minimum distance, between lines \ref{qrf:init} and \ref{qrf:fin} of Algorithm \ref{alg:qrf}. In this case, the scalar products $\vect{r_{SC}} \cdot \vers{e}_\vect{i}$ are stored in line \ref{qrf:store} and eventually employed to compute  $d=d^2_{min}$ in the following loop.

\begin{algorithm}[ht!]
\caption{- QRI}
\begin{algorithmic}[1]
\Function{QRI}{$\vect{r_{SC}}, R, \vers{e}_\vect{T}, \vers{e}_\vect{L}, \vers{e}_\vect{W}, c_T, c_L, c_W$}
\State $d \gets 0$ \Comment{initialising minimum distance}
\For{$i \in \{T,L,W\}$}
\State $a \gets \vect{r_{SC}} \cdot \vers{e}_\vect{i}$
    \If{($(l \gets a + c_i) < 0$)}
        \State \textbf{if} ($l < -r$) \textbf{return} $false$ \Comment{quick rejection test}  \label{qri:out1}   
        \State $d \gets d + l^2$
    \ElsIf{($(l \gets a - c_i) > 0$)}
        \State \textbf{if} ($l > r$) \textbf{return} $false$ \Comment{quick rejection test}  \label{qri:out2}    
        \State $d \gets d + l^2$
    \EndIf
\EndFor
\State \textbf{if} ($d \leq r^2$) \textbf{return} $true$ \Comment{checking overlap}
\State \textbf{return} $false$
\EndFunction
\end{algorithmic}
\label{alg:qri}
\end{algorithm}

\begin{algorithm}[ht!]
\caption{- QRF}
\begin{algorithmic}[1]
\Function{QRF}{$\vect{r_{SC}}, R, \vers{e}_\vect{T}, \vers{e}_\vect{L}, \vers{e}_\vect{W}, c_T, c_L, c_W$}
\State $d \gets 0$ \Comment{initialising minimum distance}
\For{$i \in \{T,L,W\}$} \label{qrf:init}
    \State $a_{i} \gets \vect{r_{SC}} \cdot \vers{e}_\vect{i}$ \label{qrf:store}
    \State \textbf{if} {($a_{i}<-c_i-R$ \textbf{or}
    \Statex \hspace{40pt} $a_{i}>c_i+R$)} \textbf{return} $false$ \Comment{quick rejection test}
\EndFor \label{qrf:fin}
\For{$i \in \{T,L,W\}$}
    \If{($a_{i} < - c_i$)}  \label{qrf:sum1}
        \State $l \gets a_{i} + c_i$
        \State $d \gets d + l^2$
    \ElsIf{($a_{i} > c_i$)} \label{qrf:sum2}
        \State $l \gets a_{i} - c_i$
        \State $d \gets d + l^2$
    \EndIf
\EndFor
\State \textbf{if} ($d \leq R^2$) \textbf{return} $true$ \Comment{checking overlap}
\State \textbf{return} $false$
\EndFunction
\end{algorithmic}
\label{alg:qrf}
\end{algorithm}

\noindent The different location of the quick rejection tests in QRI and QRF is expected to determine a difference in the efficiency of the two algorithms, which is analysed in detail in Section 4. Additionally, QRI and QRF quick rejection tests depend on both $c_i$ and $R$, so
these algorithms' efficiency are expected to be influenced also by sphere and cuboid geometry.
Finally, keeping in mind the potential application in computational colloid science, where crowded systems are usually simulated, the efficiency of QRI and QRF is also influenced by the system packing, which determines the probability for an attempted move to produce an overlap.

A parallel version of Algorithm \ref{alg:arvo}, generalised for randomly oriented cuboids and using Streaming SIMD Extension (SSE), was also proposed \cite{Larsson2007}. SSE is an instruction set that makes use of specific CPU registers to perform simple operations in parallel \cite{Thakkar1999}. By substituting the \textit{if} statements in lines \ref{qrf:sum1} and \ref{qrf:sum2} of Algorithm \ref{alg:qrf} to compute the minimum distance, with the \textit{max} and \textit{min} functions available in SSE, the computation of the minimum distance can be vectorized. This algorithm, running in parallel and thus significantly faster than QRI and QRF, does not need quick rejection tests. A pseudocode for this algorithm, here named after the SSE instruction set, is presented in Algorithm \ref{alg:sse} for the general case of randomly oriented cuboids.

\begin{algorithm}[ht!]
\caption{- SSE}
\begin{algorithmic}[1]
\Function{SSE}{$ \vect{r_{SC}}, R, \vers{e}_\vect{T}, \vers{e}_\vect{L}, \vers{e}_\vect{W}, c_T, c_L, c_W$}
\For{$i \in \{T,L,W\}$}
    \State $a_{i} \gets \vect{r_{SC}} \cdot \vers{e}_\vect{i}$ \Comment{vectorising the dot product}
\EndFor
\For{$i \in \{T,L,W\}$} \Comment{vectorising the cycle}
        \State $l_{i} \gets \min(a_{i} + c_{i}, 0) + \max(a_{i} - c_i, 0)$
        \State $l_{i} \gets l_{i}^2$
\EndFor
\State \textbf{if} $(l_{T} + l_{L} + l_{W} \leq R^2)$ \textbf{return} $true$ \Comment{checking overlap}
\State \textbf{return} $false$
\EndFunction
\end{algorithmic}
\label{alg:sse}
\end{algorithm}

Finally, we present our own algorithm, which incorporates a number of elements providing additional efficiency when compared to Algorithms \ref{alg:arvo}, \ref{alg:qri} and \ref{alg:qrf}, and versatility when compared to Algorithm \ref{alg:sse}. A new element that significantly simplifies the algorithm is the use of the absolute value to estimate the minimum distance. In addition, we employed OpenMP directives for an SIMD parallelization of the cycle \cite{Vanderpas2017} without using SSE intrinsic instructions. The advantage of avoiding SIMD intrinsics is that one can vectorize a code written in Fortran, for which these instructions are not available. Given the heterogeneous nature of the communities using collision-detection algorithms and their preference for likely different programming languages, an user-friendly code is a crucial advantage. Our algorithm, referred to as OpenMP Cuboid Sphere Intersection (OCSI), proved to be efficient and functional for both C and Fortran 90 (F90). Its pseudocode is presented in Algorithm \ref{alg:ocsi}.

\begin{algorithm}[ht!]
\caption{- OCSI}
\begin{algorithmic}[1]
\Function{OCSI}{$\vect{r_{SC}}, R, \vers{e}_\vect{T}, \vers{e}_\vect{L}, \vers{e}_\vect{W}, c_T, c_L, c_W$}
\For{$i \in \{T,L,W\}$} \Comment{this cycle is vectorised}
    \State $l_{i} = \max( |\vect{r_{SC}} \cdot \vers{e}_\vect{i}| - c_i, 0)$
    \State $l_{i} = l_{i}^2$
\EndFor
\State \textbf{if} $(l_{T} + l_{L} + l_{W} \leq R^2)$  \textbf{return} $true$ \Comment{checking ovlerlap}
\State \textbf{return} $false$
\EndFunction
\end{algorithmic}
\label{alg:ocsi}
\end{algorithm}

\section{Computational Details}
To test the relative performance of the above  algorithms, we have developed two versions of the same program in C and in F90 that detect collision between one cuboid and one sphere. The dimensions of the cuboid and sphere are given in units of the cuboid thickness $T$, which is our unit length, and do not change within the same detection-collision test. In particular, the colloid length and width are $L^* \equiv L/T$ and width $W^* \equiv W/T$, respectively, whereas the sphere radius is $R^* \equiv R/T$. For each of the cuboid shapes analysed, we have pondered the impact on the algorithms' efficiency of changing the sphere radius between $R^*=0.05$ and $R^*=5$. To control the value of the acceptance ratio, i.e. the percentage of random configurations that do not produce overlaps, the sphere $\mathcal{S}$ is generated within a spherocuboid. This spherocuboid, centred and oriented as $\mathcal{C}$, results from the Minkowski addition \cite{Schneider1993} of $\mathcal{C}$ and a sphere larger than $\mathcal{S}$ and whose diameter is optimized to obtain the desired acceptance rate. A dedicated program optimises the volume of the spherocuboid according to the target value of the acceptance ratio and the dimensions of both $\mathcal{C}$ and $\mathcal{S}$, which are specified as input parameters. To generate a configuration, $\mathcal{C}$ is initially aligned with the reference axes and its centre is set as origin, while the centre of $\mathcal{S}$ is randomly positioned within the volume of the spherocuboid. Then, the reference system is randomly rotated and the cuboid-sphere overlap checked. For consistency, the section of the code that calls the overlap function is the same as that proposed by Larsson \textit{et al} \cite{Larsson2007}. The time spent by each algorithm to detect collisions is computed for 3 independent sets of $N_c=2\times 10^6$ configurations and then averaged out. The efficiency of the algorithms has been assessed on Intel® Core™ i5-8500 CPU @ 3.00GHz (Coffee Lake) CPU using Intel Compilers. In this work, configurations of cuboids with $L^*=[1, 20]$, $W^* =[1, 20]$ and spheres with $R^* = \{0.05,0.5,5\}$ have been tested. 


\section{Results and discussion}

\begin{figure*}[t!]
    \centering
    \includegraphics[width=\textwidth]{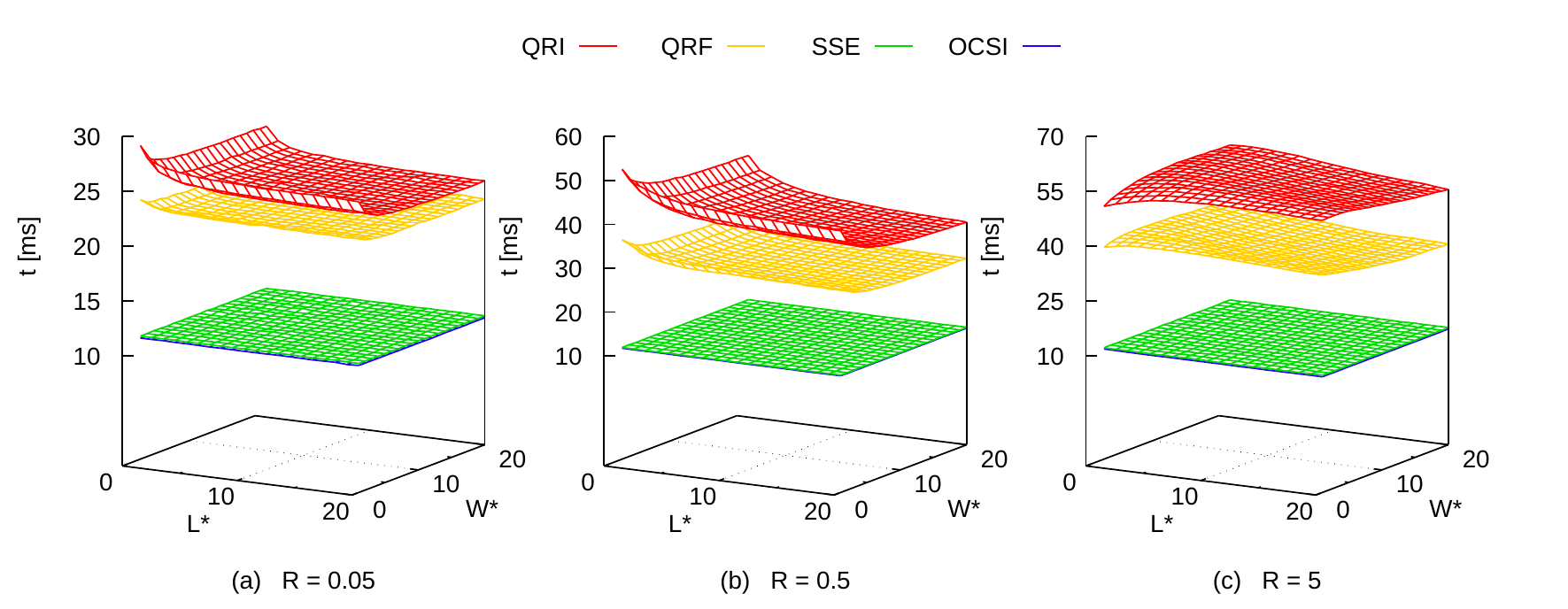}
    \caption{Run-times of algorithms written in C that detect collision between one cuboid of length $L^*$ and width $W^*$ and one sphere of radius $R^*=0.05$ (\textit{a}), 0.5 (\textit{b}) and 5 (\textit{c}). Each test generates $2\times 10^6$ random configurations at constant acceptance ratio of 40\%.}
    \label{fig:C_multi}
\end{figure*}

\begin{figure*}[t!]
    \centering
    \includegraphics[width=\textwidth]{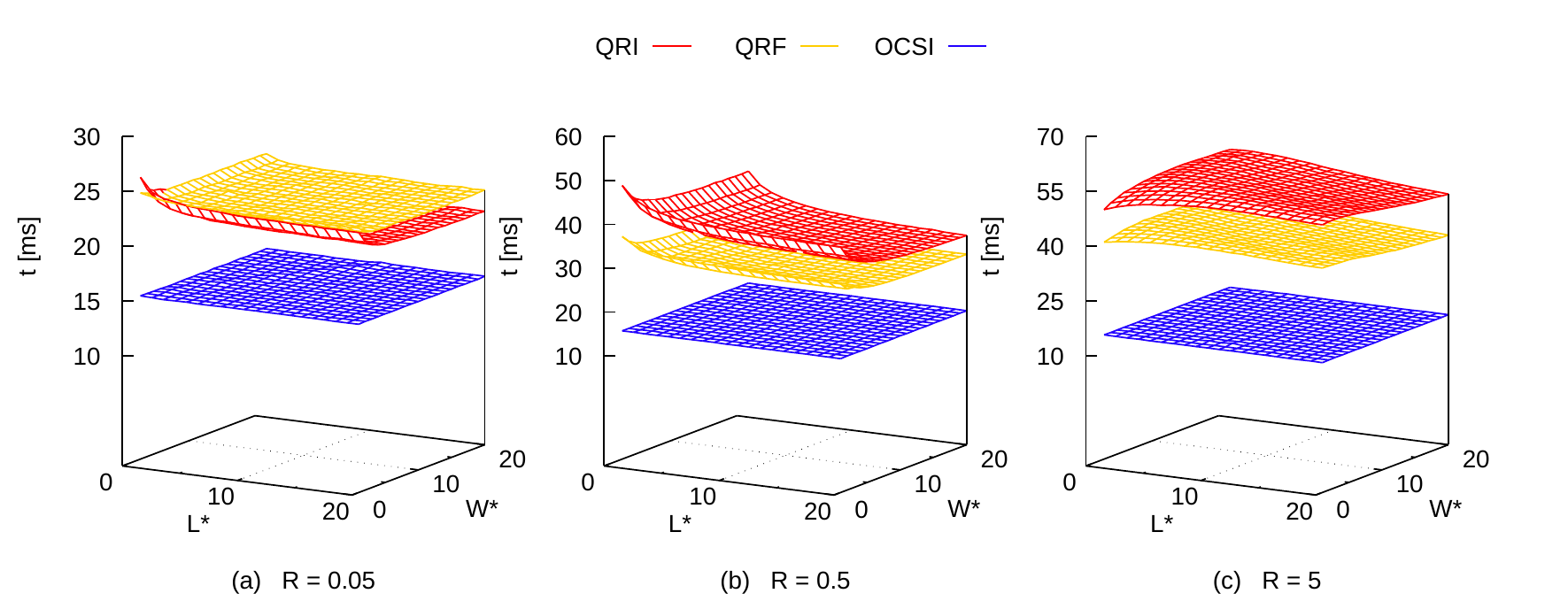}
    \caption{Run-times of algorithms written in F90 that detect collision between one cuboid of length $L^*$ and width $W^*$ and one sphere of radius $R^*=0.05$ (\textit{a}), 0.5 (\textit{b}) and 5 (\textit{c}). Each test generates $2\times 10^6$ random configurations at constant acceptance ratio of 40\%.}
    \label{fig:FOR_multi}
\end{figure*}

The relative performance of each algorithm is assessed in Fig.\ \ref{fig:C_multi} and Fig.\ \ref{fig:FOR_multi} for codes written in C and F90, respectively. Fig.\ \ref{fig:C_multi} offers a benchmark between QRI, QRF, SSE and OCSI, while Fig.\ \ref{fig:FOR_multi} only for QRI, QRF and OCSI, being the SSE algorithm unsuitable for Fortran. Both figures report the run-times for detecting collisions between one cuboid of $1 \le L^* \le 20$ and $1 \le W^* \le 20$ and one sphere of radius $R^*=0.05$ (frames \textit{a}), 0.5 (frames \textit{b}) and 5 (frames \textit{c}). The $N_c$ random configurations tested per run have been produced at the constant acceptance ratio of 40\%, which is within the usual range of values  employed in Metropolis Monte Carlo simulations of hard-core particles \cite{Frenkel1996}. It is evident that SSE and OCSI perform significantly better than QRI and QRF under the conditions specified here, although we have also tested cuboids of larger length and width (up to $100T$) with the same acceptance ratio and observed very similar tendencies. The difference in performance is especially evident at $R^*=5$ as SSE and OCSI run-times are up to 5 and 6 times faster than QRF and QRI, respectively. In general, C codes show a better performance than F90 codes, although this difference is not substantial. Interestingly enough, SSE and OCSI do not present any relevant dependence on the cuboid and sphere geometry, being the run-times basically constant across the whole range of dimensions. This is probably due to the SIMD parallelism implemented, different from QRI and QRF, which have to run in serial for their use of quick rejection tests (see lines 6 and 9 in QRI and line 5 in QRF). Basically, if the quick rejection test is true for the first dot product, the algorithms exit the loop with negative result ($\mathcal{C}$ and $\mathcal{S}$ do not overlap) with no need to compute the remaining two. 

The geometry of both cuboid and sphere exhibits a very intriguing effect on the performance of QRI and QRF as the shape of the run-time surfaces in the $L^*W^*$ plane suggests. More specifically, for spheres with $R^*=0.5$ (frames \textit{b} in Figs.\ \ref{fig:C_multi} and \ref{fig:FOR_multi}) the time required for the collision detection decreases upon increasing the cuboid dimensions, with the shortest time detected at $L^*=W^*=20$ (disk-like cuboid). Larger spheres, with $R^*=5$ (frames \textit{c} in Figs.\ \ref{fig:C_multi} and \ref{fig:FOR_multi}), induce a different performance resulting in an opposite concavity of the run-time surface as compared to that observed for smaller spheres. In this case, the slowest detection is measured at $(L^*, W^*)=(4, 8)$ and $(3, 4)$ for QRI and QRF in C program, respectively, and $(3, 9)$ for both QRI and QRF in F90 program. In general, QRF is faster than QRI, except when the spheres are especially small ($R^*=0.05$) and F90 is used. The difference in performance between QRI and QRF might be due not only to how data are stored and read by C and F90 compilers, but also to the value set for the acceptance ratio. As a matter of fact, Larsson and coworkers had already noticed that the run-times of QRI and QRF were very similar for acceptance ratios of approximately 50\%, although their collision-detection method 
tests run-times for sets of configurations with cuboids and sphere of random dimensions \cite{Larsson2007}.

To more easily compare the efficiency of the algorithms tested, the run-times reported in the frames of Figs.\ \ref{fig:C_multi} and \ref{fig:FOR_multi} have been averaged out for each value of $R^*$. The resulting average run-times are reported in Tables \ref{tab:cav} and \ref{tab:forav}. We stress that these average run-times should be taken as indicative values for QRI and QRF as their speed strongly depends on the cuboid geometry. We observe that QRI and QRF average run-times tend to be longer for larger spheres, with no significant difference between C and F90. By contrast, both SSE and OCSI are completely insensible to the sphere radius as no relevant change in their average run-time is detected between $R^*=0.05$ and 0.5.

\ctable[
 label = tab:cav,
 pos = ht!,
 caption = {Average run-times of the C version of  algorithms that detect collisions between one cuboid of $1 \le L^* \le 20$ and $1 \le W^* \le 20$ and one sphere of radius $R^*$ over $2\times 10^6$ configurations with 40\% of acceptance ratio.},
 width = \columnwidth,
]{KKKKK}{
}{
                                             \FL
 $R^*$  & QRI       & QRF      & OCSI & SSE  \NN
        & (ms)      & (ms)     & (ms) & (ms) \ML
   0.05 & 24.7      & 22.6     & 11.5 & 11.7 \NN
   0.5  & 39.4      & 29.1     & 11.5 & 11.7 \NN
   5.0  & 53.9      & 38.2     & 11.6 & 12.0 \LL
}

\ctable[
 label = tab:forav,
 pos = ht!,
 caption = {Average run-times of the F90 version of algorithms that detect collisions between one cuboid of $1 \le L^* \le 20$ and $1 \le W^* \le 20$ and one sphere of radius $R^*$ over $2\times 10^6$ configurations with 40\% of acceptance ratio.},
 width = \columnwidth,
]{KKKK}{
}{
                                      \FL
 $R^*$  & QRI       & QRF      & OCSI \NN
        & (ms)      & (ms)     & (ms) \ML
   0.05 & 22.0      & 23.4     & 15.3 \NN
   0.5  & 36.0      & 29.9     & 15.5 \NN
   5.0  & 52.8      & 40.8     & 15.5 \LL
}

We also notice that the performance of OCSI and SSE is very similar, with our algorithm between 0.2 and 0.4 ms faster than SSE. This difference, \textit{per se}, would not be especially significant if the overlap checks were limited to $2\times 10^6$ configurations. However, the typical number of configurations generated in Monte Carlo simulations of colloids is usually a few millions per particle, which are rarely less than a few thousands. Moreover, because colloids can be especially dense systems, one random configuration might generate more than a single collision. Consequently, it is reasonable to assume that, within a single Monte Carlo simulation, a collision-detection algorithm might be called between $10^3$ and $10^5$ times the configurations explored here. This would produce a difference of no more than a few tens of seconds between OCSI and SSE, which is still irrelevant. The main advantage of using OCSI is that it is based on automatic vectorization and employs OpenMP libraries to be parallelized, making it a very user-friendly algorithm. Also for this reason, OCSI can be directly implemented in Fortran, with no need to compile mixed C/Fortran codes. 


\section{Conclusions}
In summary, we have benchmarked four different collision-detection algorithms that check the occurrence of overlaps between one cuboid and one sphere. Our analysis focused on a specific acceptance ratio, which is within the usual range applied to efficiently sample the configuration space of hard-core systems in Monte Carlo simulations \cite{Frenkel1996}. We notice that SSE has been previously tested for different acceptance ratios and did not show relevant changes in performance \cite{Larsson2007}. A similar tendency is also expected for OCSI, but should be confirmed by further tests. While QRI and QRF are observed to be geometry-dependent, SSE and OCSI are basically insensible to the cuboid anisotropy and sphere radius and, thanks to automatic vectorization, they are also significantly faster. In particular, the OCSI algorithm proved to be especially valuable in terms of performance and simplicity of implementation in both C and F90. It should be stressed that the method applied to generate the sphere around the cuboid is crucial to provide a robust comparison between different algorithms. The choice of the spherocuboid as a sampling volume allows to precisely set the desired acceptance ratio and guarantees that the algorithms are tested for all the possible positions of the sphere around or inside the cuboid. This is especially relevant to fairly assess the performance of QRI and QRF, due to their use of quick rejection tests. In Monte Carlo simulations, where the generation of configurations follows a different procedure, the performance of collision-detection algorithms, most likely affected by the degree of system order and packing, should be tested. Finally, it is important to mention that the OCSI algorithm allows for the calculation of the cuboid-sphere minimum distance, hence suggesting future study to determine a suitable interaction potential beyond mere hard-core interactions. The formal proof reported here can also be useful to test the intersection of cuboids with particles of different shape.


\section{Acknowledgements}
The authors acknowledge financial support from the Leverhulme Trust Research Project Grant RPG-2018-415, Dr Benedetto Di Paola (University of Palermo) for a critical reading of the manuscript, and the use of Computational Shared Facility at the University of Manchester. 

\section{Data availability}
The source code of the program for the optimisation of the spherocuboid volume, the C and F90 versions of the benchmark program and the raw data required to reproduce these findings are available to download from http://dx.doi.org/10.17632/w7g3ynkc6n.1.


\bibliography{overlap}



\appendix

\section{On the minimum distance between a sphere and a randomly oriented cuboid}
\label{app:obbsphmin}

Let $\mathbf{V}$ be a $n$-dimensional vector space in an orthonormal basis $\mathcal{B}=\big\{ \vers{x}_\vect{1},\vers{x}_\vect{2},...,\vers{x}_\vect{n} \ \big| \ \vers{x}_\vect{i} \cdot \vers{x}_\vect{j} =\delta_{ij} \big\}$, with $\delta_{ij}$ the Kronecker delta. 
The set of points of a cuboid $\mathcal{C}$ in $\mathbf{V}$ is

\begin{equation}
\label{eq:def_cub}
\vect{C}=\vect{r_C}+\sum_{i=1}^n \alpha_{i}c_{i}\vers{e}_\vect{i},
\end{equation}

\noindent where $\vect{r_C}$ is the position of the centre of the cuboid, $c_i>0$ is a scalar equal to half of the cuboid length, width or thickness, $\alpha_i$ is also a scalar with values in $[-1, 1]$, and $\vers{e}_\vect{i}$ is a unit vector that defines the orientation of $\mathcal{C}$. Specifically $\vers{e}_\vect{i} \cdot \vers{e}_\vect{j}=\delta_{ij}$, so also $\mathcal{B'}=\{ \vers{e}_\vect{1},\vers{e}_\vect{2},...,\vers{e}_\vect{n} \}$ is an orthonormal basis for $\mathbf{V}$.
The minimum distance between $\vect{C}$ and a random point $\vect{r_S}$ reads

\begin{equation}
\label{eq:def_min}
\min\Bigg(\bigg\|\vect{r_S}-\vect{C}\Bigg\|\Bigg)=\min\Bigg(\bigg\| \vect{r_S}-\vect{r_C}-\sum_{i=1}^n \alpha_{i}c_{i}\vers{e}_\vect{i} \Bigg\| \Bigg).
\end{equation}

\noindent 
Since $\mathcal{B'}$ is an orthonormal basis for $\mathbf{V}$, the vector $\vect{r_{SC}}=\vect{r_S}-\vect{r_C}$ can be written as

\begin{equation}
\label{eq:rpc_redif}
\vect{r_{SC}} = \sum_{i=1}^n \big( \vect{r_{SC}} \cdot \vers{e}_\vect{i} \big)\ \vers{e}_\vect{i}
\end{equation}

\noindent and substituting Eq.~\ref{eq:rpc_redif} in Eq.~\ref{eq:def_min}

\begin{equation}
\label{eq:min_sum}    
\begin{gathered}
\min\Bigg(\bigg\| \sum_{i=1}^n \Big\{ \big(\vect{r_{SC}} \cdot \vers{e}_\vect{i} \big) - \alpha_{i}c_{i} \Big\}\ \vers{e}_\vect{i} \bigg\| \Bigg)= \\
=\sqrt{\min \Bigg(\sum_{i=1}^n \sum_{j=1}^n \Big\{ \big(\vect{r_{SC}} \cdot \vers{e}_\vect{i} \big) - \alpha_{i}c_{i} \Big\} \Big\{ \big(\vect{r_{SC}} \cdot \vers{e}_\vect{j} \big) - \alpha_{j}c_{j} \Big\}\ \vers{e}_\vect{i} \cdot \vers{e}_\vect{j} \Bigg) }= \\
=\sqrt{ \sum_{i=1}^n \min \Bigg( \Big\{ \big(\vect{r_{SC}} \cdot \vers{e}_\vect{i} \big) - \alpha_{i}c_{i} \Big\}^2 \Bigg) }
\end{gathered}
\end{equation}

\noindent The last term in Eq.~\ref{eq:min_sum} has been obtained considering that $\vers{e}_i \cdot \vers{e}_j=\delta_{ij}$ and that every member of the sum depends on just one value $\alpha_i$, hence they are all independent. It is sufficient to calculate only one term of this sum as all dimensions are equivalent. In particular, this term equals zero if the following conditions are met:




\begin{equation}
\label{eq:zero}
\alpha_{i}c_{i} - \big(\vect{r_{SC}} \cdot \vers{e}_i \big) = 0 \ \Leftrightarrow \ \alpha_{i} = \frac{\vect{r_{SC}} \cdot \vers{e}_i}{c_i}
\end{equation}

\noindent Because $\alpha_i=[-1,1]$, this implies that 

\begin{equation}
\label{eq:zero2}
    -1 \le \frac{\vect{r_{SC}} \cdot \vers{e}_i}{c_i} \le 1 \ \Leftrightarrow \ \Bigg|\frac{\vect{r_{SC}} \cdot \vers{e}_i}{c_i} \Bigg| \le 1 \ \Leftrightarrow \ \big|\vect{r_{SC}} \cdot \vers{e}_i\big| \le c_i
\end{equation}

\noindent If $|\vect{r_{SC}} \cdot \vers{e}_i| > c_i$, then $\left(\vect{r_{SC}} \cdot \vers{e}_i \right) > c_i$ or $\left(\vect{r_{SC}} \cdot \vers{e}_i\right) < -c_i$. The former inequality implies that


\begin{equation}
\label{eq:case1}
\begin{split}
\min\Bigg( \Big\{\alpha_{i}c_{i} - \big(\vect{r_{SC}} \cdot \vers{e}_i \big) \Big\}^2 \Bigg) &= \\
= \Big\{c_{i} - \big(\vect{r_{SC}} \cdot \vers{e}_i \big) \Big\}^2 &= \Big\{c_{i} - \big|\vect{r_{SC}} \cdot \vers{e}_i\big| \Big\}^2
\end{split}
\end{equation}

\noindent while, the latter inequality implies

\begin{equation}
\label{eq:case2}
\begin{split}
\min\Bigg( \Big\{\alpha_{i}c_{i} - \big(\vect{r_{SC}} \cdot \vers{e}_i \big) \Big\}^2 \Bigg) &=\\= 
\Big\{- c_{i} - \big(\vect{r_{SC}} \cdot \vers{e}_i \big) \Big\}^2 &= \Big\{- c_{i} + \big|\vect{r_{SC}} \cdot \vers{e}_i\big| \Big\}^2 =\\ &=\Big\{c_{i} - \big|\vect{r_{SC}} \cdot \vers{e}_i\big| \Big\}^2
\end{split}
\end{equation}

\noindent Because the solution of Eqs. \ref{eq:case1} is the same as that of \ref{eq:case2}, if $|\vect{r_{SC}} \cdot \vers{e}_i| > c_i$, then one can write

\begin{equation}
\label{eq:case_fin}
\begin{split}
\min\Bigg( \Big\{\alpha_{i}c_{i} - \big(\vect{r_{SC}} \cdot \vers{e}_i \big) \Big\}^2 \Bigg) &=\\= \Big\{c_{i} - \big|\vect{r_{SC}} \cdot \vers{e}_i\big| \Big\}^2 &= 
\Big\{\big|\vect{r_{SC}} \cdot \vers{e}_i\big| -c_{i} \Big\}^2
\end{split}
\end{equation}

\noindent The solutions of Eqs.\ \ref{eq:zero} and \ref{eq:case_fin} can be incorporated in a single equation by using a step function defined as

\begin{equation}
\label{eq:def_steps}
\Theta(x) = \bigg\{
\begin{matrix}
0 & x \le 0  \\
1 & x > 0 
\end{matrix}
\end{equation}

\noindent Therefore, the minimum distance of a point $\vect{r_S}$ from the surface of a cuboid $\mathcal{C}$ reads

\begin{equation}
\label{eq:fine}
\min\Bigg(\bigg\|\vect{r_S}-\vect{C}\bigg\|\Bigg)=\sqrt{ \sum_{i=1}^n \Theta \Big(\big|\vect{r_{SC}} \cdot \vers{e}_i\big| -c_{i}  \Big) \Big\{ \big|\vect{r_{SC}} \cdot \vers{e}_i\big| -c_{i}  \Big\}^2 }
\end{equation}

\noindent Finally, a cuboid $\mathcal{C}$ overlaps with a sphere of radius $R$ and centre in $\vect{r_S}$, if the following inequality is satisfied:

\begin{equation}
\label{eq:over}
\sqrt{ \sum_{i=1}^n \Theta \Big(\big |\vect{r_{SC}} \cdot \vers{e}_i\big| -c_{i}  \Big) \Big\{ \big|\vect{r_{SC}} \cdot \vers{e}_i\big| -c_{i}  \Big\}^2 } \le R
\end{equation}

\end{document}